# Spectrographs with holographic gratings on freeform surfaces: design approach and application for the LUVOIR mission

Eduard R. Muslimov[a,b*], Marc Ferrari[a], Emmanuel Hugot[a], Jean-Claude Bouret[a], Coralie Neiner[c], Simona Lombardo[a], Gerard Lemaitre[a], Robert Grange[a];

[a]Aix Marseille Univ, CNRS, CNES, LAM, Marseille, France; [b]Kazan National Research Technical University named after A.N. Tupolev –KAI, 10 K. Marx, Kazan 420111, Russia; [c]Observatoire de Meudon, LESIA, 14/5, place Jules Janssen, Meudon Cedex, 92195 France

**ABSTRACT**

In the present paper we demonstrate the approach to use a holographic grating on a freeform surface for advanced spectrographs design. On the example POLLUX spectropolarimeter medium-UV channel we chow that such a grating can operate as a cross-disperser and a camera mirror at the same time. It provides the image quality high enough to reach the spectral resolving power of 126 359-133 106 between 11.5 and 195 nm, which is higher than the requirement. Also we show a possibility to use a similar element working in transmission to build an unobscured double-Schmidt spectrograph. The spectral resolving power reaches 2750 for a long slit. It is also shown that the parameters of both the gratings are feasible with the current technologies.

**Keywords:** freeform optics, holographic grating, spectral resolution, LUVOIR mission, optical design.

## 1. INTRODUCTION

Freeform optics is a rapidly emerging direction in the optical technology, which is expanding the borders of achievable optical systems performance. With use of such surfaces it becomes possible to create a wide-field fast optics with fewer number of optical elements and also create new systems with geometry, which was impossible with ordinary aspheres.

Application of a freeform optical surfaces for design of diffractive optical elements represents an attractive prospect. Combining of a complex surface with a special grooves pattern could allow to get more degrees of freedom for aberrations correction and achieve a qualitatively new level in spectroscopy, imaging and beam shaping.

Such a possibility was recently considered and demonstrated by a number of authors. For example, an optical design with a blazed grating on a non-symmetrical tilted elliptical surface was developed and fabricated for the IGIS spectrograph [1]. However, in that case the grating had equally spaced straight grooves (in projection to the tangent plane). Another freeform grating was used in the Offner-type design for the ELOIS spectrometer[2]. It had a complex freeform shape without rotational symmetry and the grooves pattern varying in order to maintain their positioning with respect to the local surface normal. In general, application of freeform surfaces for imaging spectrographs design was considered in different studies[3,4]. One of such publications[5] demonstrates an Offner-Chrisp system with a grating on a general freeform surface described by Zernike fringe polynomials. Also there are ongoing research activities, which will allow to combine a freeform surface, complex grooves pattern and blazed grooves profile in a single optical element. Finally, a number of publications have shown the possibilities of application of freeform diffractive elements in adjacent fields like display technologies, or image reconstruction[7,8].

In the present paper we consider the most general case, when a grating with curved and unequally spaced grooves is imposed over a freeform surface. It is supposed that the grating is manufactured holographically, i.e.it represents a recording of interference pattern from two coherent sources. We developed a set of tools for description of such holographic freeform gratings in Zemax. The primary application of these tools was the design of POLLUX spectropolarimeter for LUVOIR[9] mission. And we show that using of the freeform grating allows to obtain the required image quality with the minimal number of optical surfaces. However, the possible applications are not limited by this particular case. It is demonstrated that a transmission freeform grating can help to achieve a high performance of a double-Schmidt type spectrograph.

---

[*] eduard.muslimov@lam.fr; phone +33 4 91 05 69 18; lam.fr

Thus the paper is organized as follows: chapter 2 presents the ways used to describe the freeform surface shape and the grating pattern; chapter 3 demonstrates usage of such an optical element for the POLLUX instrument design with estimation of the obtained image quality; chapter 4 presents a demonstrative optical design with a transmission freeform grating with the performance assessment and section 5 contains the general conclusions on the study.

## 2. DESIGN APPROACH

To describe a holographic grating on a freeform surface one need to define the surface sag and the normals and then to set up a procedure of diffraction computation in any point of this surface. Below we describe in details the means, used to model both of the points practically in Zemax.

### 2.1 Freeform description

The freeform surface in our case is described by one of the orthonormal polynomials sets – Zernike or Legendre polynomials.

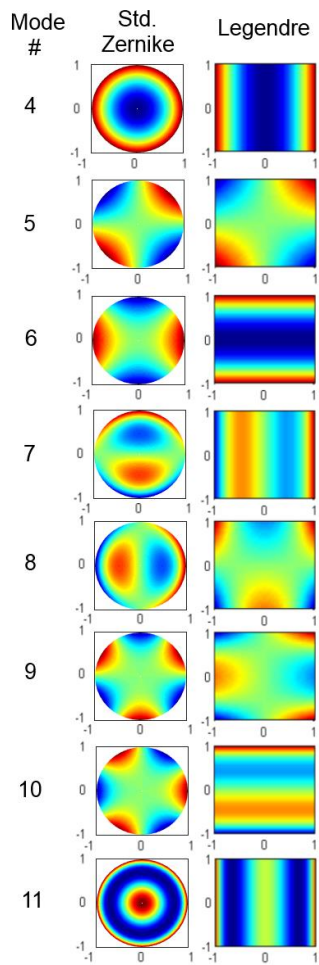

Figure 1. Modes of the orthonormal polynomials used for the freeform optical surfaces description.

The Zernike polynomials[10] are widely used in optical engineering. They are orthonormal over the unite circle, what simplifies the freeforms design and optimization with them. Another polynomial set chosen for this task is the Legendre polynomials. They are orthonormal over the unite square. Advantages of their use in some cases were demonstrated in a number of publications[11,12]. To facilitate the further explanations we provide sag diagrams for the 4th to 11th modes of each set on Figure 1 (the piston, tip and tilt are the same, so they are excluded). It is necessary to mention that a freeform surface can be described any of these polynomial sets, but due to some features of the numerical optimization process one way of the freeform description can be preferable[13].

## 2.2 Grating description

For modelling of a holographic grating we used the general raytracing procedure[14]. The vectorial equation used for computations is:

$$\vec{N} \times (\vec{r_i} - \vec{r_d}) = k \frac{\lambda}{\lambda_0} \vec{N} \times (\vec{r_1} - \vec{r_2}) \qquad (1)$$

Here $\vec{N}$ is the normal vector, $\vec{r_i}, \vec{r_d}$ are vectors of the recording rays, $\vec{r_1}, \vec{r_2}$ are that of the incident and diffracted rays, $k$ is the order of diffraction, $\lambda_0$ and $\lambda$ are the recording and working wavelengths respectively.

Hereafter we assume that the grating is recorded by two point sources. So the recording and operating geometry is presented on Figure 2. In this case the grating works in reflection, but for a transmission grating the equation and all the definitions remain similar except of the sign before the diffracted ray vector and inclusion of the refraction index into equation (1).

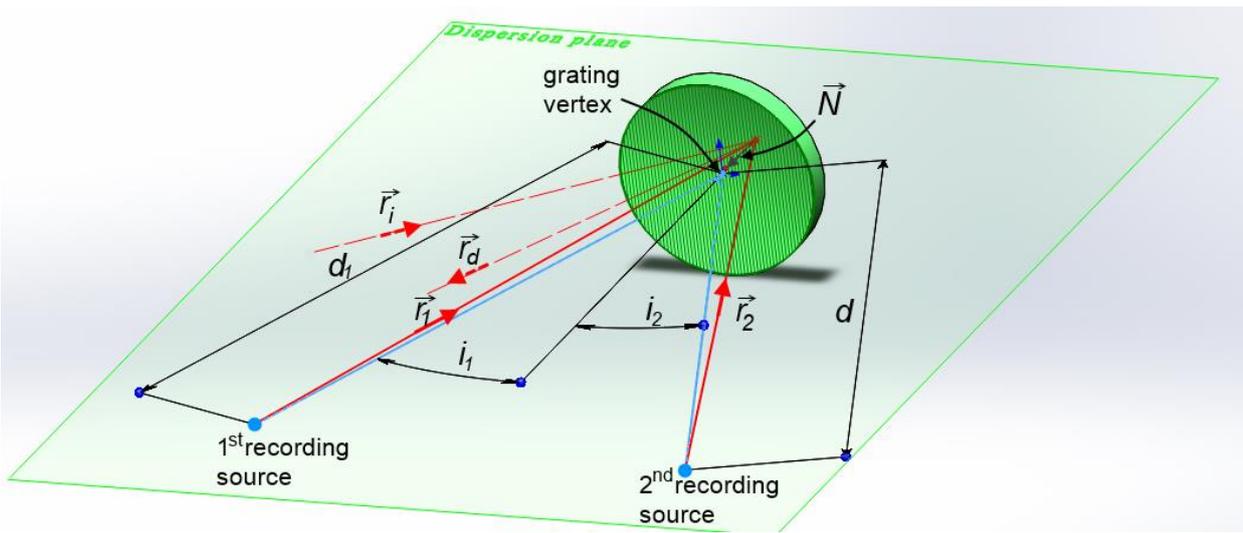

Figure 2. Definition of the holographic grating recording and operation geometry.

Combining the given descriptions of surface shape and grooves pattern one can create a user defined surface in an optical design software like Zemax (using customized .dll library) and after that use the standard raytracing and optimization tools.

## 3. OPTICAL DESIGN WITH A REFLECTION GRATING

### 3.1 Design description

The primary application of the developed design tools is optical design for POLLUX – a high resolution UV spectropolarimeter for LUVOIR mission. The target spectral resolving power of the instrument is 120 000 in the extended UV region 90-390 nm[15]. Such a high resolution can be achieved in an instrument with 3 channels, when each of them represents an echelle spectrograph. It was decided to split the channels as follows: far UV (FUV) channel operates from 90 to 124.5 nm and is fed by a flip mirror; medium UV (MUV) channel works between 118.5 and 195 nm, while near UV (NUV) channel works in the range 195-390 nm and they are separated by a dichroic splitter. It must be also mentioned, that the spectral length of a diffraction order should be 6 nm at least to avoid splitting of some broadened analytical lines.

The design solutions of the polarimetric units are different for each channel, while the spectral part is similar for all the 3 channels. The MUV channel optics general view is shown as an example on Figure 3. The entrance beam at F/20 passed through the pinhole and the polarimeter is collimated by an off-axis parabolic (OAP) mirror. Then it is dispersed by the echelle grating. The cross-disperser represents a freeform holographic grating, so it simultaneously separates the echelle's diffraction orders and focuses them on the detector. Use of such a complex element allows to correct the aberrations and achieve a high resolution and also decrease the total number of bounces and thus increase the throughput.

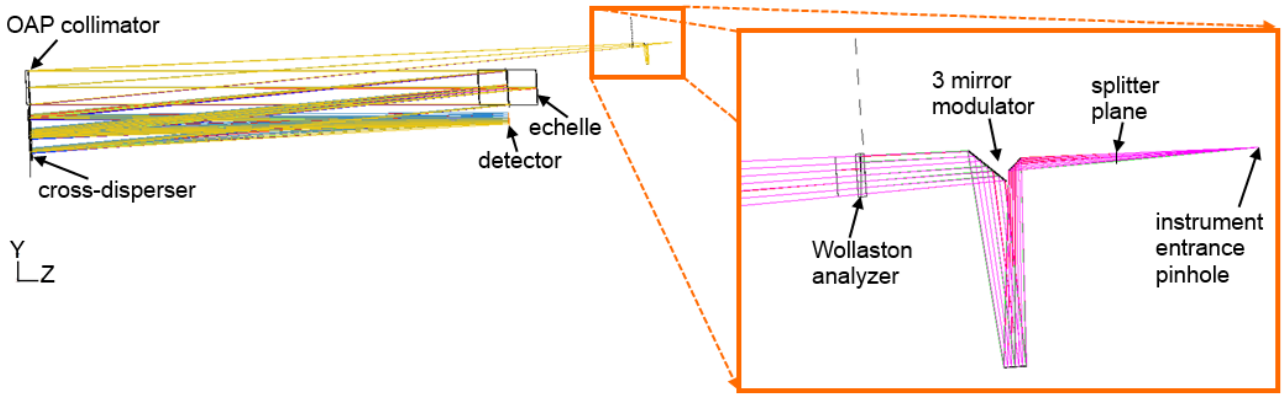

Figure 3. Optical design of the POLLUX instrument MUV channel:

left – general view, right – zoom-in of the polarimetric unit.

One of the main features of this design is spatial separation of the spectral components on the cross-disperser's surface (see Figure 4, left). It means that the aberrations for this element considerably vary across the working range. So the grating properties and the surface shape should vary locally to compensate the aberrations. It also requires use of a freeform holographic grating. After the design and optimization the grating frequency equals to 212.3 1/mm and the recording sources rectangular coordinates are (100.194,1913.946) and (-99.558,1936.898) in the case of use of *Ar* laser. The focal length of the grating acting as a camera mirror is 1200 mm and its clear aperture is 215.4x98.3 mm. The surface shape is described by the vertex sphere and 6 Zernike terms (primary coma, astigmatism and trefoil, 3$^{rd}$ order spherical, astigmatism and quadrofoil). The deviation from the best-fit sphere (BFS) is shown on Figure 4, right as a colormap. The root-mean square (RMS) deviation is 2.27 μm and the peak-to-valley (PTV) value is 3.36 μm.

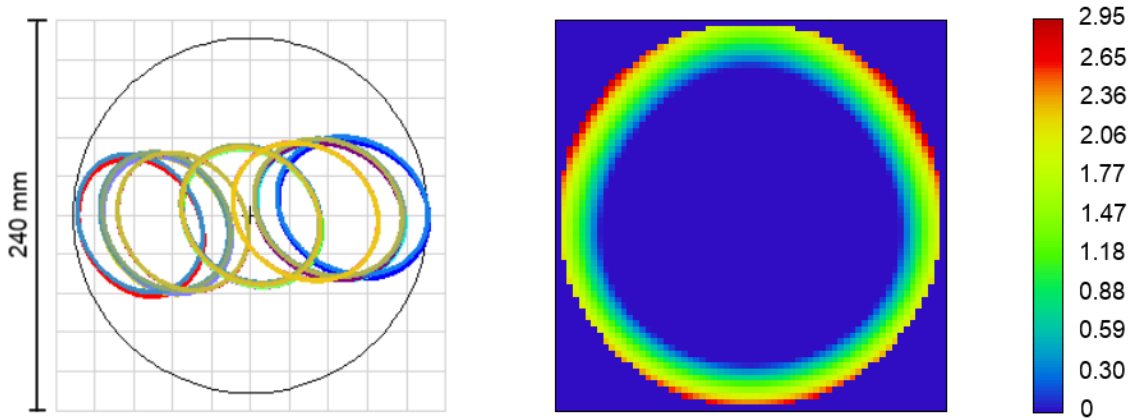

Figure 4. POLLUX MUV cross-disperser grating surface:

left – footprint diagram, right – map of the deviation from BFS in microns.

### 3.2 Performance analysis

During the design and optimization process the image quality is assessed via the spot diagrams (see Figure 5 for an example of diagrams for a single order). It is an easy way to compute an image quality estimation. One should keep in mind that the minimized function is an RMS size of the spot and that the X (horizontal) direction corresponds to the echelle dispersion direction, so the X size of the spot defines the spectral resolution. Also it should be noted that the order spectral length is 3.7 nm without the part overlapping with the adjacent orders, though the full length on the detector corresponds to 6.1 nm.

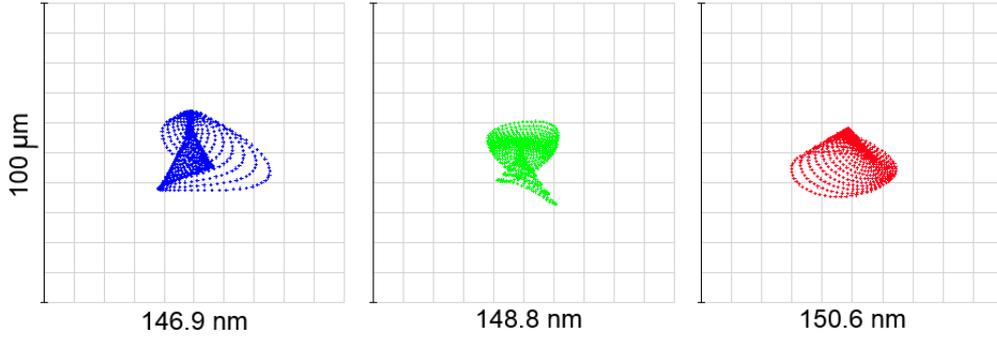

Figure 5. Sample spot diagrams of the POLLUX MUVchannel in the 40th order.

However, the dimensions of the spots don't correspond to the spectral resolving power directly. In order to compute the spectral resolving power the spectrograph's instrument functions (IF) are calculated. The entrance slit width is 31.2 μm, what corresponds to 0.03" on sky. The results for the same control wavelengths in a single order are given on Figure 6.

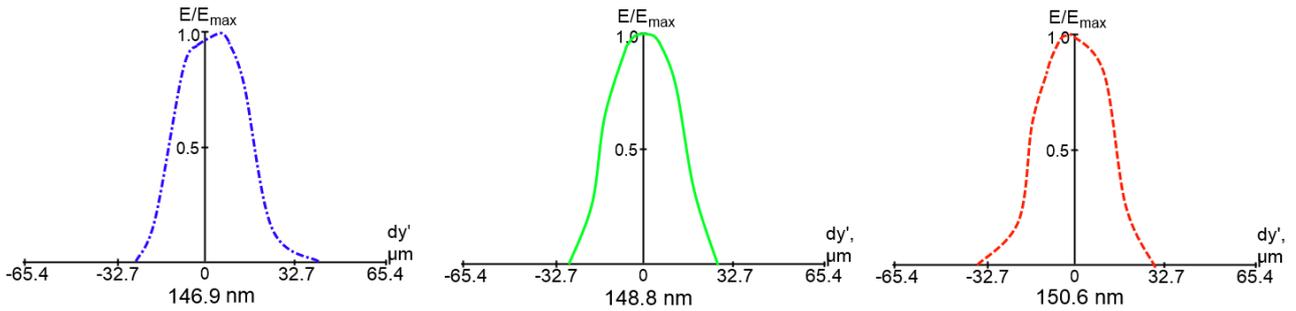

Figure 6. Sample instrument functions of the POLLUX MUVchannel in the 40th order (the entrance slit projection width is 31.2 μm).

As one can see, the aberrations correction is good enough, so the FWHM (full width at half maximum) of IF equals to the slit width. The same data re-computed to spectral resolving poser units is summarized in Table 1.

Table 1. POLLUX MUV spectral resolving power summary.

| Wavelength, nm | Order | Reciprocal linear dispersion, nm/mm | Instrument function FWHM, μm | Spectral resolving power |
|---|---|---|---|---|
| 188.9 | 31 | 0.047 | 31.2 | 128 909 |
| 191.9 | | | 31.2 | 131 008 |
| 195.0 | | | 31.95 | 133 106 |
| 150.6 | 40 | 0.036 | 31.2 | 132 673 |
| 148.8 | | | 31.2 | 131 035 |
| 146.9 | | | 31.95 | 126 359 |
| 120.2 | 50 | 0,029 | 31.95 | 129 700 |
| 119.0 | | | 31.2 | 131 502 |
| 118.0 | | | 31.2 | 130 391 |

The spectral resolving power is higher than the required value by ~10%, that allows us to leave some margins for possible manufacturing errors and misalignments.

In general, this modeling shows that with a freeform holographic grating it is possible to achieve the required spectrograph performance, while the grating recording scheme is feasible and the surface asphericity is small.

# 4. OPTICAL DESIGN WITH A TRANSMISSION GRATING

## 4.1 Design description

However, the design tools developed for Pollux can generalized and applied for other optical systems. Here as an example we consider a spectrograph with transmission freeform grating similar to the VIRUS instrument design[16]. The considered design works in the visible domain 350-550 nm with F/3.1 aperture. Similarly to the prototype we start with a design using inverted Schmidt telescope as the collimator and another Schmidt telescope as the camera$ the focal distances of the both units are equal to 420 mm. In our case the corrector plates for both of the Schmidt-type parts is united with the dispersive element thus forming the transmission freeform grating. Thanks to the Schmidt camera properties the spectrograph can operate with a long entrance slit up to 50 mm. Due to use of asymmetrical freeform surface it becomes possible to introduce tilts on the spherical mirrors and avoid any central obscuration in the optical scheme. The general view of the optical design is given on Figure 7. Note that two flat folding mirrors were introduced to decrease the overall dimensions. WE also should mention that in this case the focal plane is curved with radius of 290.96 mm, though a similar result can be obtained with a flat image plane and a field flattener.

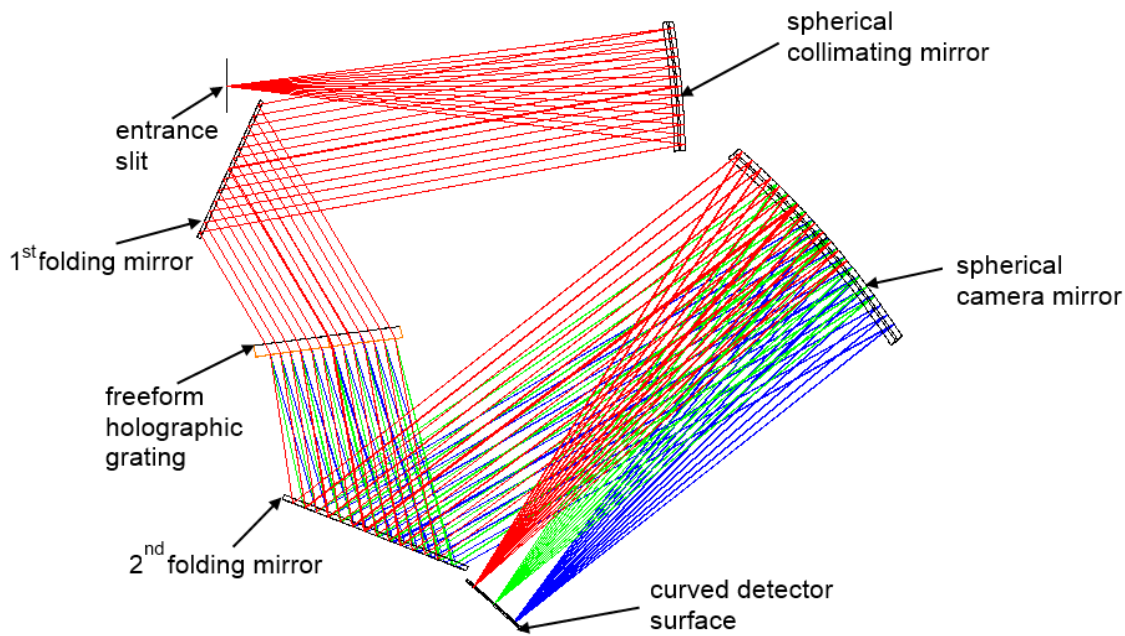

Figure 7. Optical design of a double-Schmidt spectrograph with a transmission freeform holographic grating.

The grating frequency at the vertex is 636.4 1/mm and the recording sources coordinates for a Nd:YAG laser are (403.610,2153.123) and (630.212,4027.459). Since the aperture stop is located at the grating, its clear aperture is circular with diameter of 126 mm. The optimization starts with a plane-parallel plate, but the vertex radius of curvature is set as a free parameter. For demonstrative purposes in this case we use Legendre polynomials to describe the freeform, though the surface can be also described by the Zernike modes. 9 Legendre modes, including $T_{20}$ and all the modes of the 3$^{rd}$ and 4$^{th}$ order symmetrical with respect to the *YZ* plane are used. The deviation from BFS of the surface is shown om Figure 8. The BFS radius is 6644.8mm. The RMS deviation is 70.7μm and the PTV deviation is186.6μm. These values are considerably larger than the ones obtained for previous design, however they are still manufacturable with the current level of technology. μm.

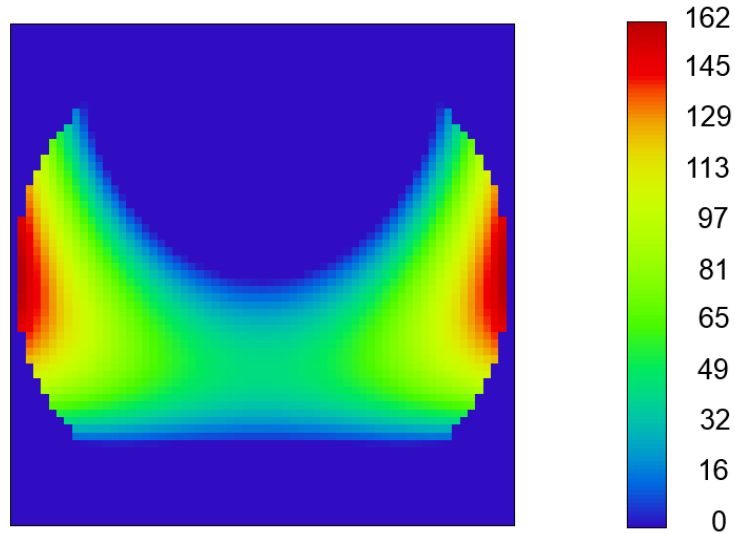

Figure 8. Deviation of the transmission freefrom grating surface from the BFS in microns.

**4.2 Performance analysis**

As well as for the first design, we consider the image quality estimations. The spot diagrams for long slit are presented on Figure 9. One can note that the aberrations are slowly growing towards the slit edges. However, the main contribution to this spot blurring is made by astigmatism. It means that the spot size in the dispersion direction remains relatively stable and the spectral resolving power doesn't change.

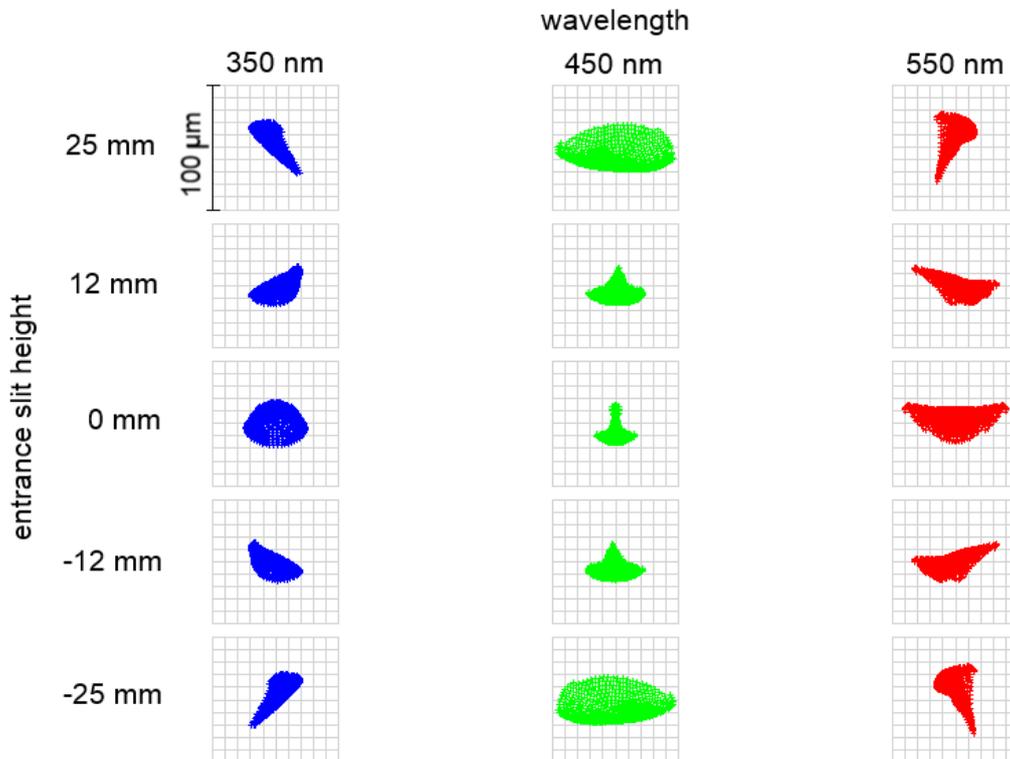

Figure 9. Spot diagrams of the double-Schmidt spectrograph for a long slit.

In order to prove and quantify the image quality we consider thee spectrograph's IF's. The IF's for the slit center are shown on Figure 10. The entrance slit width is equal to 50 μm in this case. The IF plots indicates on presence of a residual coma aberration, but the FWHM is equal to the slit width across the entire working spectral range and for all the slit points.

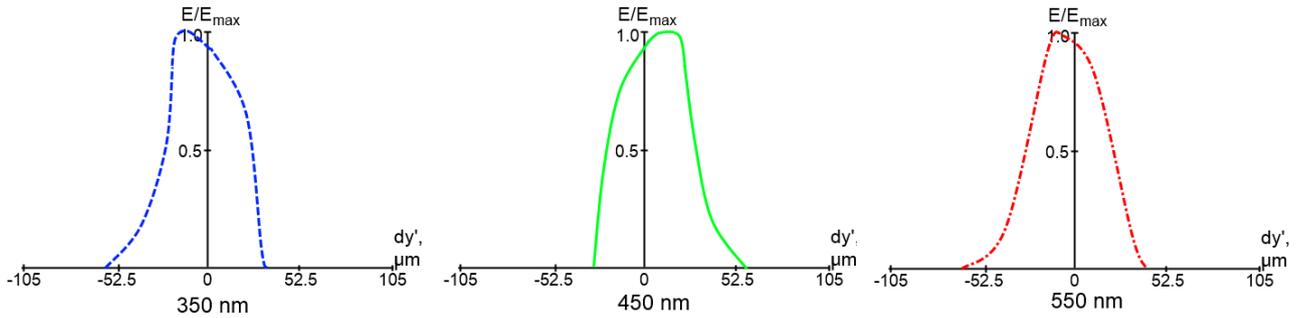

Figure 10. Instrument functions of the double-Schmidt spectrograph (the entrance slit width is 50 μm).

Table 2 summarizes the spectral resolving power data. Though a direct comparison is note fully correct, we can say that the achieved values are higher than that of the prototype design.

Table 2. The double-Schmidt spectrograph spectral resolving power summary.

| Wavelength. nm | Slit height, mm | Reciprocal linear dispersion, nm/mm | Instrument function FWHM, μm | Spectral resolving power |
|---|---|---|---|---|
| 350 | 0 | 4.0 | 50.0 | 1750 |
| 450 | | | 50.0 | 2250 |
| 550 | | | 50.0 | 2750 |
| 350 | 25 | 4.0 | 50.0 | 1750 |
| 450 | | | 50.0 | 2250 |
| 550 | | | 50.0 | 2750 |

## 5. CONCLUSIONS

In the present paper we demonstrated application of freeform holographic gratings for spectrographs design. Such an optical element provides an outstanding possibility for aberration correction, which allows to improve the performance and decrease the number of optical elements. The first application of this element is the optical scheme of POLLUX spectropolarimeter for LUVOIR mission. The freeform reflection grating works as the cross-disperser and camera simultaneously and allows to fulfill the requirements for spectral resolution and minimize the number of bounces in the scheme. The second example is a demonstrative design of a double-Schmidt spectrograph with transmission grating. The design is notable for absence of central obscuration and relatively high spectral resolving power.

Obviously, the used geometries and polynomials sets can be used in different combinations, e.g. to describe a reflective holographic grating on a Zernike-type freeform surface. This design tool and the corresponding approach to optical schemes development can be of special interest for the upcoming and future instruments intended for scientific research, especially for the astronomical instrumentation[17,18].

## ACKNOWLEDGEMENTS

The authors acknowledge the support from the European Research council through the H2020 - ERCSTG-2015 - 678777 ICARUS program. This research was partially supported by the HARMONI instrument consortium.

# REFERENCES

[1] C. Bourgenot, D. J. Robertson, D. Stelter, and S. Eikenberry, "Towards freeform curved blazed gratings using diamond machining," Proc. SPIE 9912, 99123M (2016).

[2] C. De Clercq, V. Moreau, J.-F. Jamoye, A. Zuccaro Marchi, and P. Gloesener, "ELOIS: an innovative spectrometer design using a free-form grating," Proc. SPIE 9626, 96261O (2015).

[3] C. Liu, C. Straif, T. Flügel-Paul, U. D. Zeitner, and Herbert Gross, "Comparison of hyperspectral imaging spectrometer designs and the improvement of system performance with freeform surfaces," Appl. Opt. 56, 6894-6901 (2017).

[4] L. Wei, L. Feng, J. Zhou, J. Jing, and Y. Li, "Optical design of Offner-Chrisp imaging spectrometer with freeform surfaces," Proc. SPIE 10021, 100211P (2016).

[5] J. Reimers, K. P. Thompson, J. Troutman, J. D. Owen, A. Bauer, J. C. Papa, K. Whiteaker, D. Yates, M. Farsad, P. Marasco, M. Davies, and J. P. Rolland, "Increased Compactness of an Imaging Spectrometer Enabled by Freeform Surfaces," in Optical Design and Fabrication 2017 (Freeform, IODC, OFT), OSA Technical Digest (online), paper JW2C.5 (2017).

[6] A. Z. Marchi, and B. Borguet, "Freeform Grating Spectrometers For Hyperspectral Space Applications: Status of ESA Programs," in Optical Design and Fabrication 2017 (Freeform, IODC, OFT), OSA Technical Digest (online), paper JTh2B.5 (2017).

[7] Z.Liu, Y. Pang, C. Pan, and Z. Huang, "Design of a uniform-illumination binocular waveguide display with diffraction gratings and freeform optics," Opt. Express 25, 30720-30731 (2017)

[8] P. Liu, J. Liu, X. Li, Q. Gao, T. Zhao, and X. Duan, "Design and fabrication of DOEs on multi- freeform surfaces via complex amplitude modulation," Opt. Express 25, 30061-30072 (2017)

[9] M. R. Bolcar, et al., "The Large UV/Optical/Infrared Surveyor (LUVOIR): Decadal Mission concept design update," Proc. SPIE 10398, 1039809 (2017)

[10] V. Lakshminarayanan, and A. Fleck, "Zernike polynomials: a guide," Journal of Modern Optics, 58(7), 545-561(2011).

[11] J. Ye, Z. Gao, S. Wang, J. Cheng, W. Wang, and W. Sun, "Comparative assessment of orthogonal polynomials for wavefront reconstruction over the square aperture," J. Opt. Soc. Am. A 31(10), 2304–2311 (2014).

[12] J. Ye, L. Chen, X. Li, Q. Yuan, and Z. Gao, "Review of optical freeform surface representation technique and its application," Opt. Eng. 56(11), 110901 (2017).

[13] E. Muslimov, E. Hugot, W. Jahn, S. Vives, M. Ferrari, B. Chambion, D. Henry, and C. Gaschet, "Combining freeform optics and curved detectors for wide field imaging: a polynomial approach over squared aperture," Opt. Express 25, 14598-14610 (2017).

[14] W.T. Welford, "A vector raytracing equation for hologram lenses of arbitrary shape," Optics ommunications 14(3), 322-323 (1975).

[15] E.R. Muslimov, S. Vives, E. Hugot, J.-C. Bouret, M. Ferrari, "Optical design of UV echelle spectrograph for a next generation space mission," EOS Topical Meeting on Diffractive Optics, p. 49-50 (2017).

[16] H. Lee, G. J. Hill, J. L. Marshall, B. L. Vattiat, D. L. DePoy, "Visible Integral-field Replicable Unit Spectrograph (VIRUS) optical tolerance," Proc. SPIE 7735, Ground-based and Airborne Instrumentation for Astronomy III, 77353X (2010).

[17] N. A. Thatte, et al., "The E-ELT first light spectrograph HARMONI: capabilities and modes," Proc. SPIE 9908, 99081X (2016).

[18] W. Saunders, "Efficient and affordable catadioptric spectrograph designs for 4MOST and Hector," Proc. SPIE 9147, 914760 (2014).